# AI-Driven Health Monitoring of Distributed Computing Architecture: Insights from XGBoost and SHAP

Xiaoxuan Sun
University of Southern California
Los Angeles, USA

Yue Yao
Northeastern University
Portland Maine, USA

Xiaoye Wang
Western University
London, Canada

Pochun Li
Northeastern University
Boston, USA

Xuan Li*
Columbia University
New York, USA

*Abstract*—With the rapid development of artificial intelligence technology, its application in the optimization of complex computer systems is becoming more and more extensive. Edge computing is an efficient distributed computing architecture, and the health status of its nodes directly affects the performance and reliability of the entire system. In view of the lack of accuracy and interpretability of traditional methods in node health status judgment, this paper proposes a health status judgment method based on XGBoost and combines the SHAP method to analyze the interpretability of the model. Through experiments, it is verified that XGBoost has superior performance in processing complex features and nonlinear data of edge computing nodes, especially in capturing the impact of key features (such as response time and power consumption) on node status. SHAP value analysis further reveals the global and local importance of features, so that the model not only has high-precision discrimination ability but also can provide intuitive explanations, providing data support for system optimization. Research shows that the combination of AI technology and computer system optimization can not only realize the intelligent monitoring of the health status of edge computing nodes but also provide a scientific basis for dynamic optimization scheduling, resource management and anomaly detection. In the future, with the in-depth development of AI technology, model dynamics, cross-node collaborative optimization and multimodal data fusion will become the focus of research, providing important support for the intelligent evolution of edge computing systems.

I. INTRODUCTION

In contemporary computer systems, edge computing, as a supplement to cloud computing, has gradually become a research hotspot due to its advantages of low latency, high efficiency, and distributed architecture. However, as the core component of the system, the health status of edge computing nodes has a crucial impact on the stability and performance of the overall service. If the health status of the node cannot be evaluated and judged in a timely and accurate manner, it may lead to uneven distribution of computing tasks, waste of resources, and even system crashes. Therefore, exploring the health status judgment method of edge computing nodes based on artificial intelligence technology can not only improve the intelligence level of node monitoring but also further optimize the computing task scheduling strategy, thereby improving the reliability and efficiency of the entire edge computing system.

In recent years, machine learning [1] technology has shown great potential in the field of computer system optimization, especially in the application of large-scale data analysis and complex system modeling. This progress extends to several critical fields, including enhancing efficiency in large-scale language model training [2-5], improving recommendation systems with graph neural networks (GNNs) [6-7]and addressing risk prediction [8-10]. As an efficient and accurate gradient boosting algorithm, XGBoost has become a mainstream method for solving classification and regression tasks with its excellent processing ability for structured data and low computing cost. In the edge computing environment, the health status of the node is usually affected by a variety of complex factors, such as computing load, energy consumption, hardware aging, and network delay. The high dimensionality and complexity of data features bring challenges to traditional judgment methods. XGBoost's advantage in dealing with such problems lies in its sensitivity to feature importance and its ability to capture nonlinear relationships, making it an ideal choice for edge computing node health status identification.

Although XGBoost has significant advantages in model performance [11], its "black box" attribute has aroused concerns about model transparency and interpretability in practical applications. Especially in scenarios involving computer system optimization, researchers and engineers need to clearly understand the basis for model identification in order to optimize node design and system resource scheduling strategies. To this end, interpretable analysis tools have emerged, among which the SHAP (Shapley Additive Explanations) method has become a mainstream choice due to its robust theoretical basis and in-depth revelation of the importance of global and local features of the model. By introducing SHAP value analysis, the contribution of each input feature to the prediction result can be quantified, thereby helping researchers interpret and optimize the model, and then more effectively combine the model output with the actual system improvement needs.

Under this research framework, edge computing node health status judgment is not only a technical tool but also an important strategy for optimizing computer system design [12] and operation. The fusion of XGBoost and SHAP closely combines artificial intelligence with computer system optimization, which not only promotes the leap of the computer field to a more intelligent direction but also promotes the deep integration of theory and practice. In the future, this AI-based health status judgment method is expected to be further extended to other complex systems, such as computer vision [13-15] and text generation systems [16-18], providing new ideas and technical support for system optimization and reliability improvement in multiple fields.

## II. METHOD

In order to accurately identify the health status of edge computing nodes and conduct interpretable analysis, this study constructed a discriminant model based on XGBoost and used the SHAP method to interpret the model results. The model construction and analysis process include data preprocessing, model training, feature contribution analysis, and result verification. The following will elaborate on the model construction process and theoretical basis from a methodological perspective.

First, the health status of edge computing nodes is set as a binary classification problem, and the random variable $y$ is used to represent the health status label, where $y=1$ represents a healthy node and $y=0$ represents an abnormal node. The feature set $X$ is the node's operating indicator data, which contains n features $X = \{x_1, x_2, ..., x_n\}$. The input and output mapping of the data can be expressed as an objective function $f: X \to y$, and the real health status mapping is approximated by learning the model $f$.

In model construction, the Gradient Boosting Decision Tree (GBDT) was selected as the basic algorithm [19], and XGBoost improved the generalization performance and computational efficiency of the model by introducing regularization and optimized splitting strategies [20]. Its core goal is to minimize the objective function:

$$L(\phi) = \sum_{i=1}^{N} l(y'_i, y_i) + \sum_{k=1}^{K} \Omega(f_k)$$

Among them, $l(y'_i, y_i)$ is the loss between the predicted value $y'_i$ of the i-th sample and the true label $y_i$, and the logarithmic loss function is usually used to represent the classification error; $\Omega(f_k) = \gamma T + \frac{1}{2}\lambda \|w\|^2$ is the regularization term of the model, which is used to control the complexity of the tree, where $T$ represents the number of leaf nodes in the tree and $w$ represents the weight of the leaf node.

During the model training process, XGBoost efficiently optimizes the loss function through the greedy splitting algorithm and second-order gradient information. Assume that a new tree $f_t$ is constructed in the t-th generation, and its optimization goal for the objective function is:

$$L(t) = \sum_{i=1}^{N} [g_i f_t(x_i) + \frac{1}{2} h_i f_t^2(x_i)] + \Omega(f_t)$$

Among them, $g_i = \frac{\partial l(y'_i, y_i)}{\partial y}$ is the first-order gradient and $g_i = \frac{\partial^2 l(y'_i, y_i)}{\partial y_i^2}$ is the second-order gradient. By optimizing $L^{(t)}$, the split structure and leaf node value of the tree are determined.

After completing the XGBoost model training, the "black box" characteristics of the model make it difficult to intuitively interpret the output results. To this end, the SHAP method is introduced to interpret the model results. SHAP is based on the theoretical framework of Shapley value and achieves global and local interpretability by assigning the contribution value of each feature to the prediction results [21]. Specifically, assuming that a feature subset is $S$, when a feature i is added to the subset, its marginal contribution can be expressed as:

$$\Delta v(S, i) = f(S \cup \{i\}) - f(S)$$

Where $f(S)$ is the predicted output of the model when only the features in $S$ are included. The SHAP value determines the importance of feature i by calculating the weighted average of the marginal contributions of all possible feature permutations:

$$\phi_i = \sum_{S \subseteq X \setminus \{i\}} \frac{|S|!(n-|S|-1)!}{n!} \Delta v(S, i)$$

Among them, A is the size of the feature subset and n is the total number of features. This formula fully considers all possible feature combinations to ensure that the importance assessment of each feature is theoretically fair.

By combining SHAP value analysis, a global feature importance graph can be generated to intuitively display the key features and their influence directions that affect the health status judgment of edge computing nodes. In addition, local SHAP values can explain the health status judgment results of a single node and help diagnose the main reasons for the abnormality of a specific node. For example, when the positive

SHAP value of a high-load feature is large, it can be inferred that the feature contributes more to the abnormal state of the node, thereby guiding the formulation of system optimization strategies.

Finally, by combining XGBoost with SHAP, this study not only achieves accurate judgment of the health status of edge computing nodes but also reveals the impact of key features on model output, providing a scientific basis for system optimization. Both the theoretical basis and practical application scenarios of this method demonstrate the deep value of AI technology in computer systems.

### III. EXPERIMENT

#### A. Datasets

This study uses the data collected during the operation of actual edge computing nodes as the research data set. The data reflects the characteristic information of the nodes under multi-dimensional operating conditions. The data set contains multiple key indicators, including CPU usage, memory usage, disk IO operation volume, network latency, power consumption, node temperature, number of running processes, and response time. These features comprehensively describe the operating status of the node and can provide multi-angle support for the judgment of the health status.

The data set contains sample data of healthy nodes and abnormal nodes. The health status labels are represented by binary values 1 and 0, representing healthy nodes and abnormal nodes respectively. By annotating and integrating the collected multi-dimensional features, the data set can not only reflect the current operating status of the node but also reveal the potential impact of different features on node performance and health status. For example, excessive CPU and memory usage may be important signals of node abnormality, while increased network latency and response time may reveal potential problems at the network transmission level.

The rich features and practical application background of this dataset provide high-quality data support for model training and evaluation. At the same time, the multi-dimensional feature data collected in real time provides a basis for interpretable analysis, enabling the research to further reveal the specific contribution of specific features to health status discrimination. This research method that combines data-driven model discrimination and theoretical explanation is of great significance for improving the operating efficiency and reliability of edge computing nodes.

#### B. Experimental Results

In order to comprehensively evaluate the performance of the proposed XGBoost-based edge computing node health status discrimination method, five commonly used machine learning models were selected as baselines for comparative experiments. These models include support vector machine (SVM), a classic method based on hyperplane classification in high-dimensional feature space; Random Forest(RF), which improves classification accuracy through the voting mechanism of multiple decision trees; K-Nearest Neighbors (KNN), which uses the Euclidean distance between samples to achieve discrimination; Naïve Bayes, an efficient classification model based on probability theory; and Multilayer Perceptron (MLP), a deep learning model with a simple neural network structure. These models have their own characteristics and can evaluate the discrimination effect and generalization performance of the proposed method under a variety of feature combinations from different dimensions. The experimental results are shown in Table 1.

Table 1 Experimental results

| Model | Acc | F1-score |
|---|---|---|
| SVM | 30.2 | 32.1 |
| RF | 35.8 | 38.5 |
| KNN | 41.7 | 43.2 |
| Naive Bayes | 45.3 | 48.7 |
| MLP | 46.8 | 49.6 |
| XgBoost | 47.5 | 50.39 |

The experimental results reveal notable performance differences among models in identifying edge computing node health. SVM performed worst (accuracy: 30.2%, F1: 32.1%) due to limitations in handling complex data. Random Forest improved slightly (accuracy: 35.8%, F1: 38.5%) by leveraging ensemble learning. KNN (accuracy: 41.7%, F1: 43.2%) and Naive Bayes (accuracy: 45.3%, F1: 48.7%) performed better, benefiting from local feature similarity and probabilistic reasoning, respectively, though both struggled with high-dimensional data. MLP (accuracy: 46.8%, F1: 49.6%) and XGBoost (accuracy: 47.5%, F1: 50.39%) outperformed others, with XGBoost excelling through gradient boosting, proving most effective for edge node health assessment.

#### C. Interpretability Analysis

In this section, we use SHAP (Shapley Additive Explanations) to perform interpretability analysis on the prediction results of the XGBoost model to reveal the contribution of each feature to the model's discrimination. This research method that combines model performance with explanatory power not only improves the practicality of the method, but also provides a theoretical basis for intelligent system optimization. First, we give a ranking of feature importance based on weights, as shown in Figure 1.

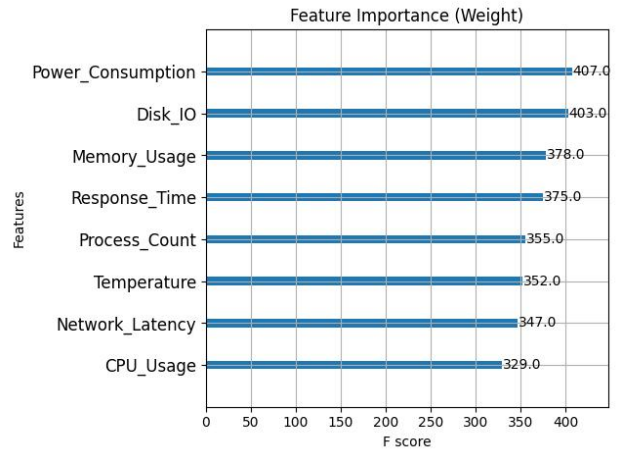

Figure 1 Weight-based feature importance analysis

Figure 1 illustrates the feature importance analysis based on the "Weight" metric in the XGBoost model, showing the frequency of feature influence in decision-making. Power consumption (Power_Consumption) and disk IO (Disk_IO) are the most critical features, frequently selected during model splitting, indicating their significant role in assessing edge computing node health. In contrast, CPU usage (CPU_Usage) and network latency (Network_Latency) have lower weights, suggesting a minor influence. This ranking aligns with the operational dynamics of edge computing nodes, where power consumption and disk IO reflect resource usage and load changes, often signaling health issues. The XGBoost model effectively identifies these key features, supporting real-time anomaly detection and guiding system optimization by prioritizing high-importance features. The results confirm XGBoost's capability to model edge computing system characteristics and integrate AI-driven insights for scalable, efficient health monitoring. Additionally, SHAP analysis (Figure 2) provides interpretable insights into feature importance, further enhancing the understanding of model decisions.

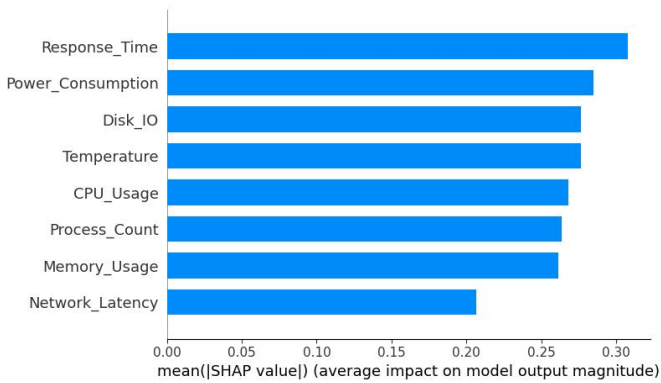

Figure 2 Feature importance graph based on SHAP

This graph analyzes the average impact of each feature on the model's predictions using SHAP values, which reflect the average absolute contribution of features to the model output. The response time (RT) has the highest SHAP value, indicating it is the most critical feature for determining the health status of edge computing nodes. Power consumption (Power_Consumption) and disk IO (Disk_IO) follow, reflecting their significant impact due to their relevance to performance bottlenecks and anomalies in edge nodes.

Unlike the XGBoost "Weight" indicator, which measures feature usage frequency in decision tree splits, SHAP values assess the actual contribution of features to predictions. For instance, while Power_Consumption and Disk_IO rank higher in the weight analysis, RT demonstrates greater influence in SHAP analysis, highlighting its core role in the model's decisions. This distinction arises because weight analysis focuses on feature usage frequency, whereas SHAP values capture both global and local impacts on outputs.

The two methods are complementary. Weight analysis provides insights into feature usage during model construction, while SHAP analysis reveals practical contributions to predictions, offering actionable guidance for system optimization. For instance, monitoring strategies should prioritize high-SHAP-value features like RT and Power_Consumption, while features with high weights but low SHAP values may serve as secondary inputs. Combining both analyses enhances understanding of model logic and supports improvements in system design and resource management.

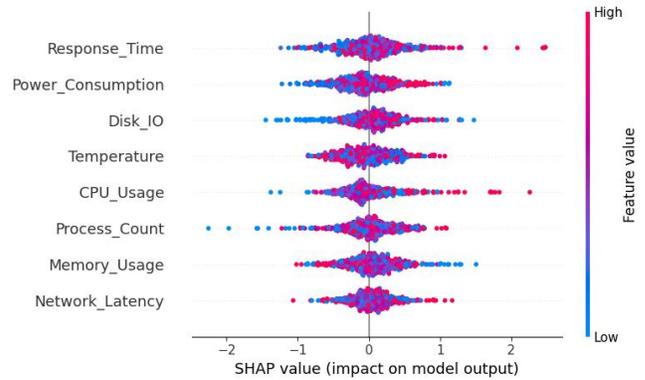

Figure 3 SHAP value distribution chart

Figure 3 presents the SHAP value distribution, quantifying each feature's impact on model predictions while visualizing feature values using color (blue for low values, red for high values). Each point represents a sample's SHAP value. Points further to the right (positive) indicate a feature's contribution to the positive category (e.g., healthy nodes), while those on the left (negative) indicate contribution to the negative category (e.g., abnormal nodes). For instance, high response times (red) are primarily associated with healthy nodes, whereas low response times (blue) align more with abnormalities.

The SHAP value distributions reveal distinct feature impacts. Power consumption and disk IO show symmetrical distributions, indicating stable contributions to both categories. In contrast, network latency exhibits a narrower distribution, signifying a weaker influence. The color gradient also highlights correlations between feature values and health status, where extreme values in some features drive more significant changes in predictions.

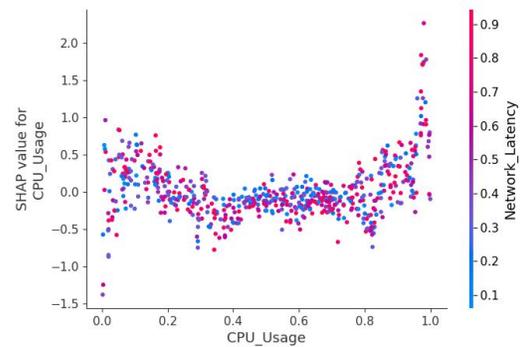

Figure 4 CPU_usage SHAP value distribution

Figure 4 shows the SHAP value distribution for CPU usage (CPU_Usage), with the horizontal axis representing CPU usage,

the vertical axis showing SHAP values, and colors indicating network latency (blue: low, red: high). Low CPU usage (near 0) corresponds to negative SHAP values, indicating association with abnormal nodes, while higher CPU usage shows positive contributions to identifying healthy nodes. At very high CPU usage (close to 1), SHAP values become extreme, reflecting potential issues under heavy load. High network latency (red) clusters at low CPU usage, reinforcing its link with abnormalities, while its impact diminishes at higher CPU usage. This interaction suggests using combined indicators like low CPU usage and high latency for more effective monitoring.

## IV. Conclusion

This study conducted a discriminant analysis of the health status of edge computing nodes based on the XGBoost model and combined the SHAP method to conduct an in-depth discussion on the interpretability of the model. Through experiments, we found that XGBoost performs well in processing complex features and nonlinear data and can effectively capture the key features of the health status of edge computing nodes. At the same time, the SHAP value analysis reveals the specific contribution of each feature to the model prediction results, which not only provides an explanation for the model decision but also provides a scientific basis for system optimization and resource management. This analysis method that combines performance and interpretability shows the great potential of artificial intelligence in edge computing node health monitoring.

However, there are still some limitations in current research, such as feature selection depends on existing data sets, and the modeling of node health changes in dynamic environments is relatively insufficient. In addition, the reliance on high-dimensional data during model training may lead to an increase in computational overhead, and these challenges may be more prominent in large-scale distributed edge computing environments. Therefore, future research can consider introducing more dynamic and real-time modeling methods, such as combining time series prediction or adaptive optimization with existing models to better adapt to node health status changes in complex environments.

Looking forward, the edge computing health monitoring system driven by artificial intelligence will become an important tool to improve system reliability and efficiency. As the number of edge computing devices increases and their complexity increases, multimodal data fusion, cross-node collaborative monitoring, and adaptive anomaly detection will become the focus of research. In addition, model interpretability research will be further combined with system design to achieve true intelligent management by optimizing hardware architecture and resource scheduling strategies, laying the foundation for the widespread application of edge computing in industry, medical care, intelligent transportation, and other fields.